\documentclass[12pt,preprint]{aastex}

\begin{document}


\shorttitle{{\it Chandra} X-ray Observations of RX J0911.4+0551}
\title{{\it Chandra} X-ray Observations of the Quadruply Lensed \\ Quasar RX J0911.4+0551}






\author{N. D. Morgan\altaffilmark{1}, 
        G. Chartas\altaffilmark{2}, 
        M. Malm\altaffilmark{1},
	M. W. Bautz\altaffilmark{1}, 
	I. Burud\altaffilmark{3},
	J. Hjorth\altaffilmark{4},
        S. E. Jones\altaffilmark{1}, 
        P. L. Schechter\altaffilmark{1}
}



\altaffiltext{1}{Department of Physics, Massachusetts Institute of
 Technology, Cambridge MA 02139; malm@space.mit.edu, mwb@space.mit.edu, 
 ndmorgan@space.mit.edu, schech@space.mit.edu, sjones@space.mit.edu}

\altaffiltext{2}{Department of Astronomy and Astrophysics, 525 Davey Laboratory,
 Pennsylvania State University, University Park, PA 16802; chartas@lonestar.astro.psu.edu}

\altaffiltext{3}{Institut d'Astrophysique et de Geophysique de Li\`ege, 
Universi\'e de Li\`ege, Avenue de Cointe 5, 4000 Li\`ege, Belgium; burud@astro.ulg.ac.be}

\altaffiltext{4}{Astronomical Observatory, University of Copenhagen, Juliane Maries Vej 30, DK-2100 Copenhagen, Denmark; jens@astro.ku.dk}


\begin{abstract}
We present results from X-ray observations of the quadruply lensed quasar 
RX~J0911.4+0551 using data obtained with the Advanced CCD Imaging Spectrometer 
(ACIS) on board the {\it Chandra X-ray Observatory}.  The 29~ks observation 
detects a total of $\sim$ 404 X-ray photons (0.3 to 7.0~keV) from the four 
images of the lensed quasar.  Deconvolution of the aspect corrected data 
resolves all four lensed images, with relative positions in good agreement 
with optical measurements.  When compared to contemporaneous optical data, one 
of the lensed images (component A3) is dimmer by a factor of $\sim$6 in X-rays with respect 
to the 2 brighter images (components A1 and A2).  Spectral fitting for the combined 
images shows significant intrinsic absorption in the soft (0.2 to 2.4~keV) 
energy band, consistent with the mini-BAL nature of this quasar, while a comparison 
with {\it ROSAT} PSPC observations from 1990 shows a drop of $\sim$6.5 in the total soft bandpass 
flux.  The observations also detect $\sim157$ X-ray photons arising from extended
emission of the nearby cluster (peaked $\sim42\arcsec$ SW of RX~J0911.4+0551) responsible for 
the large external shear present in the system.  The {\it Chandra} observation reveals the
cluster emission to be complex and non-spherical, and yields a cluster
temperature of $kT = 2.3^{+1.8}_{-0.8}$ keV and a 2.0 to 10~keV cluster luminosity 
within a 1 Mpc radius of $L_X = 7.6_{-0.2}^{+0.6} \times10^{43}$ ergs s$^{-1}$ (error bars
denote 90\% confidence limits).  Our mass estimate of the cluster within its virial radius 
is $2.3^{+1.8}_{-0.7} \times10^{14} \mbox{ M}_{\odot}$,
and is a factor of two smaller than, although consistent with, previous mass estimates 
based on the observed cluster velocity dispersion.

\end{abstract}




\section{Introduction}

Gravitational lenses that produce multiple images of quasars can be
used to measure cosmological parameters (Refsdal 1964; Kochanek 1996), 
to study the properties of the lensing galaxies (Keeton, Kochanek, \& 
Falco 1998) and to resolve structures associated with the lensed 
quasar (Bunker, Moustakas, \& Davis 2000).  For the past twenty years 
such work has been carried out at optical and radio wavelengths.  With 
the advent of the subarcsecond angular resolution of the {\it Chandra X-ray 
Observatory} (Weisskopf, O'Dell, \& van SpeyBroeck 1996), one
can hope to carry out similar studies at X-ray wavelengths.  As the
X-ray emission from quasars has different spatial and temporal
structure, X-ray studies offer a variety of new opportunities.  In the
present paper we report on {\it Chandra} observations of the quadruply
imaged quasar RX~J0911.4+0551, which have implications both for the
structure of the lensing potential and for the structure of the X-ray emitting
region.

To our knowledge, the observations of RX J0911.4+0551 presented here are 
the first X-ray observations of the system since its detection by the 
{\it ROSAT} All Sky Survey in 1990 October.  The source was optically identified 
as a gravitationally lensed $z=2.8$ quasar by Bade et al.~(1997), and later 
confirmed as a quadruple system by Burud et al.~(1998).  Lensing models
require a large external shear ($\gamma \ga 0.15$) to 
account for the system's complex image geometry.  Burud et al. (1998) first 
suggested the source of this shear to be a group of galaxies in the southwest 
vicinity of RX~J0911.4+0551, which has been spectroscopically confirmed by 
Kneib et al. (2000) to be a high-redshift cluster of at least 24 
galaxies at $\bar{z} = 0.769$.  The original {\it ROSAT} observation of 
RX~J0911.4+0551 lacked the sensitivity and angular resolution required for 
a detailed X-ray analysis of the lensed quasar and cluster.  Deconvolution of the 
{\it Chandra} observations presented here identifies all four components of 
the lensed quasar, and an adaptively smoothed image reveals the extent of the X-ray 
emission from the nearby cluster as well.  The format of the paper is as follows: we describe 
the {\it Chandra} observations and reductions of RX~J0911.4+0551 in \S 2, 
discuss X-ray astrometry and relative photometry of the lens components in \S 
2.1, and report spectral properties of the lensed quasar in 
\S 2.2.  The X-ray detection and analysis of the cluster is presented in \S 3.  We 
summarize our findings and briefly comment on possible sources of X-ray 
variability for the system in \S 4.

\section{Observations and Analysis}

RX~J0911.4+0551 was observed for $\sim$ 29~ks with the {\it Chandra X-ray 
Observatory} on 1999 November 2.  The data were obtained using the 
back-illuminated S3 chip of the Advanced CCD Imaging Spectrometer (ACIS; 
Garmire et al. 1992; Bautz et al. 1998) at a focal plane temperature of 
$-110\degr$C.  The telescope pointing placed RX~J0911.4+0551 at the default 
aim point for the ACIS-S array during the exposure.  Three periods of 
background flaring were detected, during which time the count rate reached 2 
to 4 times the nominal background rate.  Since the contamination of point 
sources from background flaring is small, these times were not filtered from 
our observation except for the cluster analysis discussed in \S 3.  
Removal of the three X-ray flares reduces the effective exposure time of the 
data to $\sim$ 26~ks.

The data were reduced using the CIAO software tools, following the science 
threads outlined on the {\it Chandra X-ray Observatory} Center (CXC) user 
support homepage.  The data used throughout this paper has
been reprocessed by the CXC using software version {\tt R4CU5UPD9}, which
results in an improved aspect solution over earlier versions of the processing 
software.  Also, in the standard CXC pipeline, the photon event positions are 
randomized by $\pm$ $0\farcs246$ (1 ACIS pixel = 0$\farcs$492), which is mainly 
to reduce aliasing effects for observations of duration less than $\sim$ 2~ks.  
Given the small angular extent of the lens ($\sim3\arcsec$), we reproduced the 
event positions using the CXC tool {\tt acis\_process\_events} without 
incorporating the randomization.

Throughout this work, we have adopted a flat cosmological model parameterized by 
$q_o = 0.5$ and $H_o = 50$ km s$^{-1}$ Mpc$^{-1}$.  At the redshift of the lensing 
galaxy ($z$ = 0.769; Kneib et al. 2000), this gives a physical scale of 8.2 kpc 
arcsec$^{-1}$.  Unless otherwise noted, error bars denote 90\% confidence limits.

\subsection{Relative Astrometry and Photometry}

In Figure 1a we show the X-ray image of RX~J0911.4+0551, resampled
at a resolution of $0\farcs1$ pixel$^{-1}$ and smoothed with 
a Gaussian of $\sigma$=$0\farcs2$.  A total of $\sim 404$ X-ray photons
in the 0.3 to 7.0~keV bandpass are estimated from the four lensed images
of the quasar, which compensates for contributions from the X-ray
background and from the nearby cluster.  Adopting the nomenclature of 
Burud~et~al.~(1998), component B is clearly resolved whereas components 
A1, A2 and A3 are not clearly separated.  To enhance the image quality we
applied the Richardson-Lucy maximum likelihood deconvolution technique 
(Richardson 1972; Lucy 1974) to the X-ray image of RX~J0911.4+0551.  The 
point spread function (PSF) incorporated in the deconvolution was generated 
with the raytrace simulator MARX v3.0 (Wise et al. 1997).  The input 
spectrum for the simulation was based on the model of the observed 
spectrum of RX~J0911.4+0551 (see \S2.2).  In particular, we chose an 
absorbed power law model with Galactic absorption of N$_{H}$ = 0.036 
$\times$ 10$^{22}$ cm$^{-2}$, intrinsic absorption at $z = 2.8$ of 
N$_{H}$ = 4.1 $\times$ 10$^{22}$ cm$^{-2}$ and a photon spectral index 
of 1.63.  The PSF and image were then sampled at a resolution of $0\farcs1$.  
In Figure 1b we show the deconvolved image of RX~J0911.4+0551 smoothed with 
a Gaussian of $\sigma$ = $0\farcs2$.  All four components are now resolved 
with relative positions in good agreement (to better than $0\farcs15$) with 
those found from optical observations.  Component B appears somewhat extended
along the EW direction in both the raw and deconvoled images, which we attribute
to the small number of counts ($\sim$ 70 in total for 0.3 to 7.0~keV) detected 
for this component.  In Table 1 we present the optical and X-ray offsets of the 
lensed images.

The ACIS PSF (when properly normalized) gives the probability distribution 
of detecting a single X-ray photon at some position with respect to the center 
of a point source.  One can therefore use several PSFs, of different relative 
intensities, to quantify the likelihood of obtaining the observed distribution
of photon counts for a given set of image flux ratios and positions (e.g., see
Cash 1979).  Specifically, we have solved for the relative flux ratios of the 
system components using the Cash (1979) {\it C} statistic to quantify the relative 
goodness of fit between the observed and predicted number of photons per bin.  
The system was modeled using four of the MARX PSFs described above; relative 
positions of the four sources were held fixed according to the optical 
astrometry listed in Table 1, but the overall position of the system was 
allowed to vary during the fit.  The best fit parameters were then determined 
by simultaneously varying the intensity of each component and the overall 
position of the system using Powell's direction-set method (Press et al. 1995) 
until a minimum {\it C} statistic was reached.  This analysis was performed 
on the raw data at a resampling size of $0\farcs06$, or an eighth of an ACIS pixel.

In Table 2 we present our optimal X-ray flux ratios for the 0.3 to 7.0~keV 
bandpass, along with contemporaneous (to within one day) $I$-band data obtained
from an ongoing monitoring campaign of RX~J0911.4+0551 using the Nordic Optical 
Telescope (NOT; Hjorth et al. 2001).  Since the separation between components A1 and A2 is 
$0\farcs48$, or slightly less than one ACIS pixel, it is prudent to discuss the combined flux 
of components A1 and A2 when comparing to the optical data.  Our maximum likelihood 
analysis yields an (A1~+~A2)/B X-ray flux ratio of 4.68 and an (A1~+~A2~+~A3)/B flux 
ratio of 4.87, which are $\sim$ 20\% and $\sim$ 30\% smaller than the respective 
optical ratios of 5.67 and 6.95.  A significant difference is observed for A3/B, 
where we find an X-ray flux ratio of 0.19 as compared to the optical ratio of 1.28.  
If we normalize the optical and X-ray fluxes at their respective (A1~+~A2) values, then 
component B is brighter by a factor of 1.2 in X-rays and component A3 is fainter by a 
factor of 5.7 in X-rays.  We briefly discuss possible implications of the faintness of 
A3 in \S 4.

\subsection{Spectral Properties}

A source spectrum of RX~J0911.4+0551 was extracted by combining events from all 
four quasar images and a background spectrum from events located within an annulus centered 
on RX~J0911.4+0551 with inner and outer radii of 15$\arcsec$ and 30$\arcsec$, 
respectively.  Redistribution matrices and ancillary response files were generated 
using the CXC tools {\tt mkrmf} and {\tt mkarf}.  A fit of an absorbed 
power-law model with intrinsic and Galactic absorption in the energy range 
0.5-6.0~keV yields a photon spectral index of 1.45$^{+0.24}_{-0.20}$ and a 
0.2-2.4~keV flux (approximating the {\it ROSAT} bandpass) of f$_{X}$ = 
3.3$^{+0.1}_{-0.1}$ $\times$ 10$^{-14}$ erg s$^{-1}$ cm$^{-2}$.  A fit of a 
simple absorbed power-law model with Galactic absorption of N$_{H}$ = 0.036 
$\times$ 10$^{22}$ cm$^{-2}$, considering only energies above 1.5~keV, yields 
a photon spectral index of 2.22$^{+0.43}_{-0.48}$, which is typical for high 
redshift radio-quiet quasars (Reimers et al. 1995). As we show in Figure 2, 
extrapolating this model to energies below 1.5~keV clearly indicates that 
considerable intrinsic absorption is present.  This low-energy absorption is 
consistent with the mini-BAL nature of RX~J0911.4+0551 (Bade et al.~1997), and 
has been observed in at least one other BAL quasar (e.g., PHL~5200, Mathur et 
al.~1995).  For RX~J0911.4+0551, the 0.2-2.4~keV flux of the unabsorbed 
model is f$_{X}$ = 1.0$^{+0.5}_{-0.4}$ $\times$ 10$^{-13}$ erg s$^{-1}$ cm$^{-2}$, and 
represents what we would expect to observe were the absorber not present.

RX~J0911.4+0551 was observed with the {\it ROSAT} PSPC on 1990 October with a 
count-rate of 0.020 $\pm$ 0.0088 cnts s$^{-1}$ (1 $\sigma$).  Assuming the simple absorbed 
power-law model from above, the corresponding X-ray flux for the {\it ROSAT} observation 
is f$_{X}$ = (2.2 $\pm$ 1.1) $\times$ 10$^{-13}$ erg s$^{-1}$ cm$^{-2}$ (1 $\sigma$).  This 
includes emission from both the gravitational lens and extended emission from the 
nearby cluster (see \S 3).  Within a circular aperture centered on RX~J0911.4+0551 of 
diameter 1$\arcmin$ (twice the half-power diameter of the {\it ROSAT} PSPC), the 
{\it Chandra} estimate of the flux due to solely cluster emission is 
(6.6 $\pm$ 1.0) $\times10^{-15}$ ergs cm$^{-2} $ s$^{-1}$ (1 $\sigma$).  Therefore, the cluster 
contribution to the original {\it ROSAT} flux was small, on the order of a few percent.  
When we compare the estimated {\it ROSAT} flux from RX~J0911.4+0551 to the {\it Chandra} observation, 
we find it has dropped by a factor of $\sim6.5 \pm 3.3$ (1 $\sigma$) over the 10-year period.  Our 
spectral fits to the {\it Chandra} spectrum of RX~J0911.4+0551 indicate that a factor of 3.0 $\pm$ 1.5 
(1 $\sigma$) variation in the 0.2-2.4~keV flux is possible if one assumes that the presently observed 
intrinsic X-ray absorber were not present during the {\it ROSAT} PSPC observation.

\section{Cluster Analysis}

The {\it Chandra} observation of RX~J0911.4+0551 provides the 
first detailed X-ray study of the nearby cluster reported 
by Kneib et al. (2000).  The extent of the X-ray emission from the cluster 
is apparent after adaptively smoothing the image with the 
CXC tool {\tt csmooth} (Ebeling, White and Rangarajan, 2000).  
Contours for the smoothed image are displayed in Figure 3, 
and are shown overlaid on an {\it IJK} composite exposure of 
the RX~J0911.4+0551 field from Burud et al. (1998).  When creating 
the smoothed image, we have used a conservative lower bound 
on the signal to noise ratio (SN) of the smoothing kernel of 
SN$\geqslant$3, and have also excluded emission from within a small 
($\sim4\arcsec$ radius) region around RX~J0911.4+0551 itself.  The 
smoothed emission profile consists of a bright, extended 
component peaked $\sim42\arcsec$ southwest of component B, 
but is clearly elongated in the direction toward 
RX~J0911.4+0551.  An iterative centroiding algorithm 
(Buote \& Canizares 1992) applied to the 
binned event positions yields a centroid for the extended 
emission of $\Delta \alpha = -12\farcs0$, $\Delta \delta = 
-39\farcs8$, where offsets are with respect to the centroid 
of component B.  The contour morphology suggests that the 
cluster mass distribution is complex and non-spherical, and 
possibly not dynamically relaxed.  If the X-ray emission 
follows the mass distribution (which, simulations suggest, 
it need not always do during merging; see, e.g. Roettiger, 
Stone \& Mushotzky 1998) then our image implies that the cluster 
is responsible for some fraction of the mass convergence at the 
location of the lens.  

Within an ellipse of 70$\arcsec$ $\times$ 90$\arcsec$ (dashed outline 
in Figure 3) encompassing the bulk of the cluster emission, we 
find a total of 157 net cluster counts above an estimated background of 
222 counts.  The corresponding 0.3 to 7.0~keV flux (assuming the 
spectral form described below) is f$_{X}$ = (2.5 $\pm$ 0.2) $\times10^{-14}$ 
erg s$^{-1}$ cm$^{-2}$ (1 $\sigma$).

To estimate an X-ray temperature for the cluster, we extracted 
a 0.3 to 7.0~keV source spectrum centered on the SW peak within 
a 20$\arcsec$ aperture, and subtracted  a background spectrum 
from an annulus with inner and outer radii of 90$\arcsec$ and 
140$\arcsec$, respectively.  Fitting the emission spectrum to an 
optically thin Raymond and Smith (1977) plasma model with a 
Galactic absorption of $N_H = 0.036 \times 10^{22} \mbox{cm}^{-2}$, 
a metal abundance of 0.3 Z$_\sun$, and a cluster redshift of 
$z=0.769$ (Kneib et al. 2000) yields a temperature of $kT = 
2.3^{+1.8}_{-0.8}$ at the 90\% confidence level.   This is 
consistent with the estimate of $kT = 4.5 \pm 1.2$ keV obtained 
by Kneib et al. (2000) from their measurement of the cluster 
velocity dispersion and the $\sigma-T_x$ relationship (Girardi 
et al. 1996).  

We do not detect enough photons to constrain the radial 
surface brightness profile of the cluster.  We obtain
a crude estimate of the integrated cluster emission by
assuming the profile is well-described by an isothermal $\beta -$model 
(Cavaliere \& Fusco-Femiano 1976) with parameters typical 
for nearby, non-cooling flow clusters ($\beta=0.6$, $r_{c}=0.22$ 
Mpc; Mohr et al. 1999).  Normalizing the model using the observed 
flux within the elliptical aperture, we predict a 2.0 to 10~keV 
luminosity within a 1 Mpc (122$\arcsec$) radius of $L_{X}=
7.6_{-0.2}^{+0.6} \times10^{43}$ ergs s$^{-1}$. 

From our temperature we may obtain a crude cluster mass estimate.  We first
estimate the virial radius for the cluster. Girardi et al. (1998) find, for
nearby clusters, that the virial radius is $R_{V} \approx 4 \sigma_{1000} 
\ h_{50}^{-1}$ Mpc, where $\sigma_{1000}$ is the radial velocity dispersion
divided by 1000 km s$^{-1}$. Assuming the virial radius scales with redshift 
as $R_{V} \sim (1+z)^{-3/2}$,
and given the observed velocity dispersion $\sigma_{1000} = 0.84$ (Kneib et 
al. 2000), we estimate $R_{V} \approx 1.4 h_{50}^{-1}$ Mpc. Assuming
an isothermal plasma with the density distribution described above 
and $kT$ = 2.3  keV, the mass within the virial radius is $M(R_{V}) =
2.3^{+1.8}_{-0.7} \times10^{14} \mbox{ M}_{\odot}$.  
Alternatively, the mass-temperature relation of Hojrth, Oukbir, \& van Kampen (1998) 
gives $M(R_V) = 3.2 ^{+2.5} _{-1.0} \times 10^{14} \mbox{ M}_{\odot}$.
Here the 90\% confidence limits are dominated by errors in the temperature,
and do not include systematic errors arising from our ignorance of the
slope of the surface brightness distribution, or from calibration uncertainties
in the mass-temperature relationship. 
Our X-ray mass estimate is somewhat smaller than, although 
consistent with, the virial estimate of 
$4.8_{-2.0}^{+2.3} \times10^{14}$ M$_{\odot}$ (1 $\sigma$; $\Omega_m = 1.0, 
\Omega_{\Lambda} = 0.0$) obtained by Kneib et 
al. (2000).  Given the asymmetry of the X-ray emission, it is reasonable to 
speculate that the X-ray gas is not in hydrostatic equilibrium. In this
case, one might well expect the emission weighted X-ray temperature to lead to an 
underestimate of the gravitating mass (e.g. Roettiger, Stone and Mushotzky 1998). 
In any event, given the large uncertainties, the agreement among the 
mass estimates must be regarded as good.

\section{Discussion and Conclusions}

Our analysis has detected two indications of X-ray variability for RX~J0911.4+0551.  
First, component A3 is dimmer in X-rays by a factor of $\sim$6 when compared to 
contemporaneous optical data.  The reason for A3's X-ray faintness is not known.
One possible explanation is differential $N_H$ column densities
along the lines of sight to the quasar images.  Although an attempt was made to search 
for differential X-ray extinction between the four lens components (the details of which 
will be presented in a forthcoming paper; Chartas et al. 2001), a preliminary analysis is 
inconclusive; the intrinsic $N_H$ column density of image A3 is poorly constrained by 
the {\it Chandra} data.  A demagnifying microlensing event of component A3 is also a 
possible, although somewhat unlikely, explanation.  A six-fold drop in flux would require 
fluctuations of the order $\Delta m \sim 2$ magnitudes; microlensing dips of this size 
are rare in theory (Witt, Mao, \& Schechter 1995).

Second, the total observed flux for the system has dropped by a factor $\sim6.5$ 
when compared to {\it ROSAT} PSPC observations obtained in 1990 October.  Such
a drop in flux may be a result of quasar engine variability.  Almaini et al. (2000) have 
observed engine variability of a factor of three in flux over a period of 
several days, so a factor of $\sim6.5$ between the {\it ROSAT} and {\it Chandra} 
observations is not unlikely.  

In conclusion, the unique imaging capabilities of {\it Chandra} has resolved 
all four images of the quadruply lensed system RX~J0911.4+0551 (detecting $\sim$404 photons 
in the 0.3 to 7.0~keV energy range), with relative image positions in good agreement with 
optical measurements.  In addition to the two indications of X-ray variability 
described above, we have also detected strong soft ($< 1.5$ keV) X-ray absorption in the 
spectra of the combined quasar images, which is consistent 
with the mini-BAL nature of the quasar.  The {\it
Chandra} observation of RX~J0911.4+0551 has also identified extended X-ray emission from 
the nearby cluster, detecting $\sim$157 photons (0.3 to 7.0 keV) in a region of $\sim$ 1.4 arcmin$^2$.
The morphology of the cluster emission suggests a complex and non-spherical
cluster mass distribution.  Our estimates of the cluster X-ray temperature ($kT=2.3^{+1.8}_{-0.8}$ keV)
and virial mass ($2.3^{+1.8}_{-0.7} \times10^{14} \mbox{ M}_{\odot}$) are
consistent with previous estimates derived from the observed velocity dispersion of the
cluster galaxies (Kneib et al. 2000).

\acknowledgments

N.D.M. and P.L.S. gratefully acknowledge support from the U.S. National Science 
Foundation through grant AST96-16866 and from NASA under contract NAS-8-37716.  
M.W.B. and M.M. acknowledge support from NASA under contracts NAS-8-37716, 
NAS-8-38252 and 1797-MIT-NA-A-38252.  J. H. acknowledges support by the 
Danish Natural Science Research Council (SNF).

\clearpage



\figcaption[fig1.ps]{Panel a: {\it Chandra} X-ray image of the
quadruply lensed quasar RX~J0911.4+0551, resampled at one-fifth of an ACIS pixel 
(1 ACIS pixel = 0$\farcs$492) and smoothed with a Gaussian of $\sigma$=$0\farcs2$.  
Panel b: Deconvolution of the Chandra X-ray image using the Richardson-Lucy algorithm.  
The deconvolved image has been smoothed with a Gaussian of $\sigma$=$0\farcs2$.  
The four components of the system are labeled according to the nomenclature of 
Burud et al. (1998).  North is up and East is left;  the width of 
each panel is $\sim 5.5\arcsec$.  See inset of Figure 3 to compare the image
configuration with an HST/NICMOS observation of the system.
\label{fig1}}

\figcaption[fig2.ps]{Top panel: {\it Chandra} spectral data of the combined 
quasar images of RX~J0911.4+0551 with a best-fit power law (including Galactic 
and intrinsic absorption) to the hard (1.5-7.0~keV) energy band.  Bottom panel:  
Deviation (in units of one standard deviation) of the observed spectral 
counts to the hard energy band fit. \label{fig2}}

\figcaption[fig3.ps]{Solid lines: X-ray contours (logarithmically spaced) 
of an adaptively smoothed {\it Chandra} image of the RX~J0911.4+0551 field 
(excluding emission from RX~J0911.4+0551 itself).
The background image is a composite optical image taken with the NOT and 
ESO/NTT (Burud et al. 1998).  X-ray contours have been aligned with the 
optical exposure according to the {\it Chandra} aspect solution.
Dashed line:  Elliptical aperture used in \S 3 to estimate the cluster 
luminosity.  Inset:  Closeup of RX~J0911.4+0551 taken from archival HST/NICMOS 
imaging of E. Falco (PI).  The 4 quasar images and lensing galaxy are labeled.
\label{fig3}}





\clearpage



\begin{deluxetable}{ccccc}
\rotate
\tablecaption{Optical and X-ray Offsets of RX J0911.4+0551 Components
\label{TABLE1}}
\tablenum{1}
\tablewidth{0pt}
\tablehead{
Telescope & \multicolumn{1}{c}{ B } & \multicolumn{1}{c}{ A1 } & \multicolumn{1}{c}{ A2 } & \multicolumn{1}{c}{ A3 } \nl
         & $\Delta \alpha$($\arcsec$), $\Delta \delta$ ($\arcsec$)                   & $\Delta \alpha$($\arcsec$), $\Delta \delta$ ($\arcsec$)& $\Delta \alpha$($\arcsec$), $\Delta \delta$ ($\arcsec$)& $\Delta \alpha$($\arcsec$), $\Delta \delta$ ($\arcsec$)}
\startdata
NOT         &  0.00, 0.00 & 2.935$\pm$0.004, -0.785$\pm$0.009 & 3.194$\pm$0.007, -0.383$\pm$0.007 & 2.922$\pm$0.008, 0.161$\pm$0.009 \nl
Chandra     &  0.00, 0.00 & 2.88$\phn$$\pm$0.20$\phn$, -0.88$\phn$$\pm$0.20$\phn$ & 3.32$\phn$$\pm$0.20$\phn$, -0.41$\phn$$\pm$0.20$\phn$ & 2.88$\phn$$\pm$0.20$\phn$, 0.17$\phn$$\pm$0.20$\phn$ \nl
\enddata

\tablecomments{Optical positions and corresponding 1 $\sigma$ error bars are from 
observations obtained with the Nordic Optical Telescope (NOT; Burud 
et al. 1998).  Relative offsets are in the sense of component A minus component B.}

\end{deluxetable}

\clearpage


\begin{deluxetable}{cccc}
\tablecaption{Contemporaneous Optical and X-ray Flux Ratios of RX J0911.4+0551 Components
\label{TABLE2}}
\tablenum{2}
\tablewidth{0pt}
\tablehead{
\colhead {Waveband} &
\colhead { (A1+A2)/B } &
\colhead { (A1+A2+A3)/B } &
\colhead { (A3)/B }}
\startdata
I band      &  5.67 $\pm$ 0.14 & 6.95 $\pm$ 0.16 & 1.28 $\pm$ 0.07 \nl
0.3-7.0 keV &  4.68 $^{+0.30}_{-0.20}$ &  4.87 $^{+0.23}_{-0.12}$ & 0.19 $^{+0.24}_{-0.17}$\nl
\enddata

\tablecomments{ $I$-band ratios and corresponding 1$\sigma$ error bars are 
from an ongoing monitoring campaign of RX~J0911.4+0551 with the Nordic
Optical Telescope (Hjorth et al. 2001).  Error bars for the X-ray ratios 
represent 90\% confidence intervals.}

\end{deluxetable}

\clearpage

\begin{figure}[h]
\vspace{7.0 truein}
\includegraphics{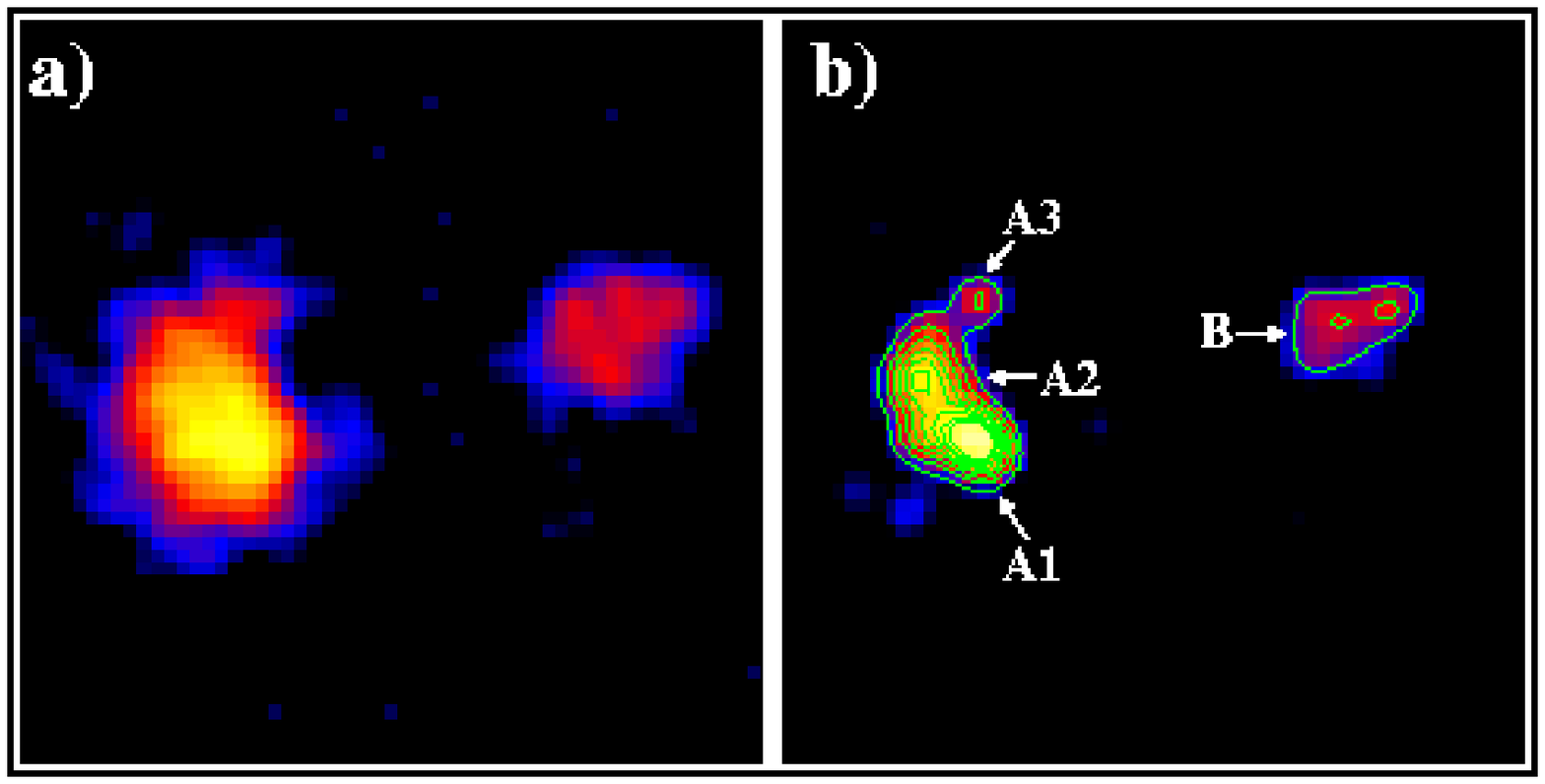}
\end{figure}
\clearpage

\thispagestyle{empty}
\begin{figure}[h]
\vspace{7.0 truein}
\includegraphics{NickMorgan.fig2.ps}
\end{figure}
\clearpage

\begin{figure}[h]
\vspace{7.0 truein}
\includegraphics{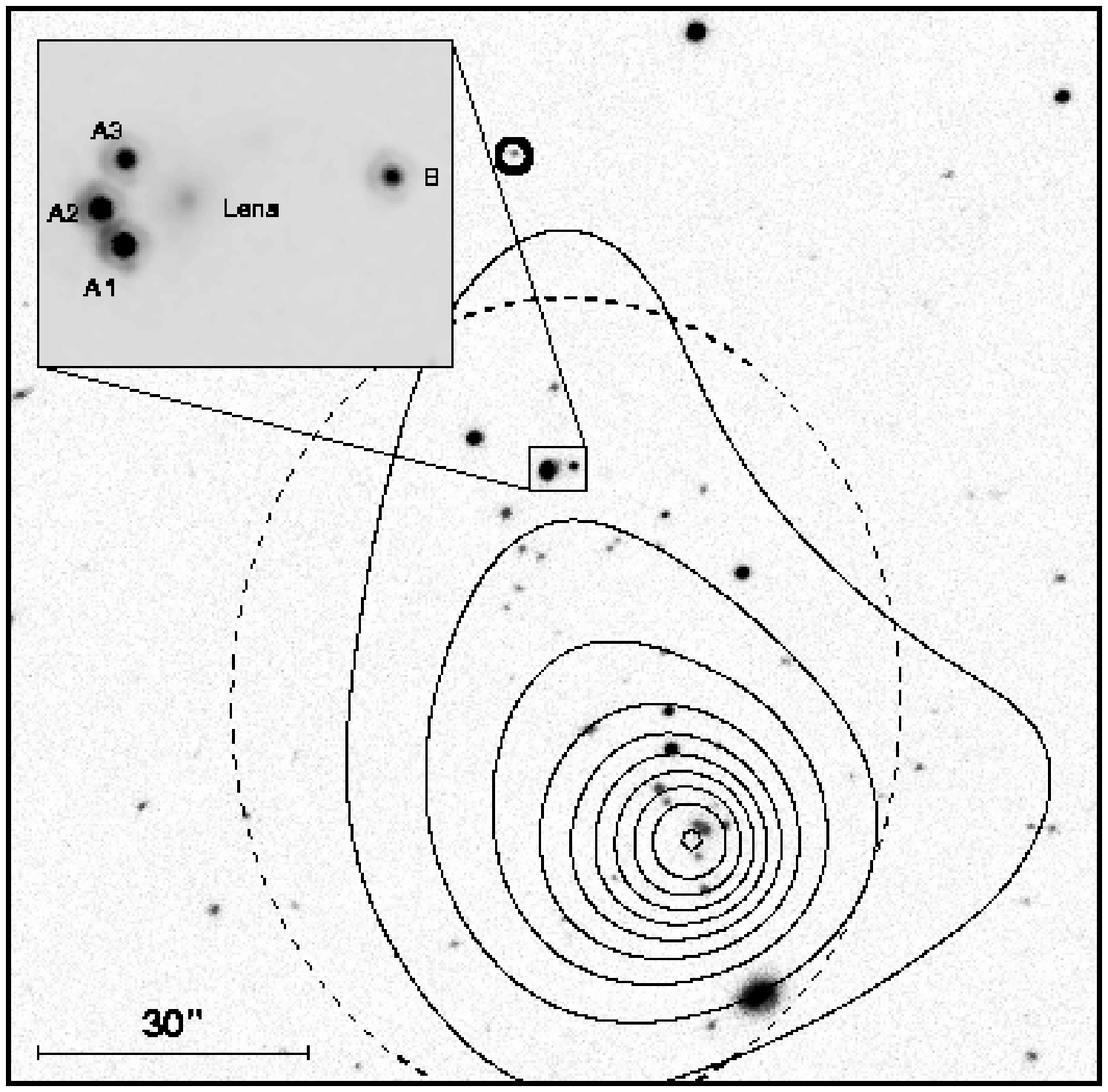}
\end{figure}
\clearpage



\begin{thebibliography}{}
\bibitem[Almaini(2000)]{2000} Almaini, O., Lawerence, A., Shanks, T., Edge, A.,
	Boyle, B. J., Georgantopoulos, I., Gunn, K. F., Stewart, G. C., \& Griffiths, R. E.
	2000, \mnras, 315, 325
\bibitem[Bade(1997)]{1997} Bade, N., Siebert, J., Lopez, S., Voges, W., \& Reimers, D. 1997,
	\aap, 317, L13
\bibitem[Bautz(1998)]{1998} Bautz, M. W., {\it et al.} 1998, X-Ray Optics, Instruments
	and Missions, ed. R. B. Hoover \& A. B. Walker, Proc. SPIE, 3444, 210
\bibitem[Bunker(2000)]{2000} Bunker, A. J., Moustakas, L. A., \& Davis, M. 2000, \apj, 531, 95
\bibitem[Buote(1992)]{1992} Buote, D. A., \& Canizares, C. R. 1992, \apj, 400, 385
\bibitem[Burud(1998)]{1998} Burud, I., Courbin, F., Lidman, C., Jaunsen, A. O., Hjorth, J., 
	{\O}stensen, R., Andersen, M. I., Clasen, J. W., Wucknitz, O., Meylan, G., \& Magain, P. 
	1998, \apj,  501, L5
\bibitem[Cash(1979)]{1979} Cash, W. 1979, \apj, 228, 939
\bibitem[Cavaliere(1976)]{1976} Cavaliere, A. \& Fusco-Femiano, R. 1976, \aap, 49, 137
\bibitem[Chartas(2001)]{2001} Chartas et al. 2001, in preparation.
\bibitem[Ebeling(2000)]{2000} Ebeling, H., White, D. A., \& Rangarajan, F. V. N. 2000,
	submitted to MNRAS. 
\bibitem[Garmire(1992)]{1992} Garmire, G. P. {\it et al.} 1992, AIAA, Space Programs and 
	Technologies Conference, Huntsville, AL, Mar 24-27
\bibitem[Girardi(1996)]{1996} Girardi, M., Fadda, D., Giuricin, G., Mardirossian, F., 
	Mezzetti, N., \& Biviano, A. 1996, \apj, 457, 61
\bibitem[Girardi(1998)]{1998} Girardi, M., Borgani, S., Giuricin, G., 
	Mardirossian, F., \& Mezzetti, M. 1998, \apj, 505, 74
\bibitem[Hjorth(1998)]{1998} Hjorth, J., Oukbir, J., \& van Kampen, E. 1998, \mnras, 298, L1
\bibitem[Hjorth(2001)]{2001} Hjorth et al. 2001, \apjl, in preparation
\bibitem[Keeton(1998)]{1998} Keeton, C. R., Kochanek, C. S., Falco, \& E. E. 1998, \apj, 509, 561
\bibitem[Kneib(2000)]{2000} Kneib, J.-P., Cohen, J. G., \& Hjorth, J. 2000, \apjl, 544, L35
\bibitem[Kochanek(1996)]{1996} Kochanek, C. S. 1996, \apj, 473, 595
\bibitem[Lucy(1974)]{1974} Lucy, L. B. 1974, \aj, 79, 745
\bibitem[Mathur(1995)]{1995} Mathur, S., Elvis, M., \& Singh, K. P. 1995, \apj, 455, L9
\bibitem[Mohr(1999)]{1999} Mohr, J. J., Mathiesen, B., \& Evrard, A. E. 1999, \apj, 517, 627
\bibitem[Press(1995)]{1995} Press, W. H., {\it et al.} 1995, {\it Numerical Recipes in C}.
	(Cambridge: Cambridge University Press)
\bibitem[Raymond(1977)]{1997} Raymond, J. C., \& Smith, B. W. 1977, \apjs, 35, 419
\bibitem[Reimers(1995)]{1995} Reimers, D., Bade, N., Schartel, N., et al., 1995, \aap, 296, L49
\bibitem[Refsdal(1964)]{1964} Refsdal, S. 1964, \mnras, 128, 307
\bibitem[Richardson(1972)]{1972} Richardson, W. H. 1972, J. Opt. Soc. Am., 62, 55
\bibitem[Roettiger(1998)]{1998} Roettiger, K., Stone, J. M., \& Mushotzky, R. F. 1998, \apj, 493, 62
\bibitem[Weisskopf(1996)]{1996} Weisskopf, M. C., O'Dell, S. L. \& van Speybroeck, L. P. 1996,
	Proc. SPIE, 2805, 2
\bibitem[Witt(1995)]{1995} Witt, H. J., Mao, S., \& Schechter, P. L. 1995, \apj, 443, 18
\bibitem[Wise(1997)]{1997} Wise, M. W., Huenemoerder, \& D. P., Davis, J. E. 2000, ADASS VI,
	A.S.P. Conf. Series, Vol. 125, 1997, ed. Gareth Hunt \& H. E. Payne, p. 477
\end{thebibliography}
\end{document}